\documentclass[11pt]{article}
\usepackage{latexsym,psfig,rotate,epsf,a4,ukleinebib3} 
 
\renewenvironment{thebibliography}[1] 
        {\begin{list}{\arabic{enumi}.} 
        {\usecounter{enumi}\setlength{\parsep}{0pt} 
         \setlength{\itemsep}{0pt} 
         \settowidth 
        {\labelwidth}{#1.}\sloppy}}{\end{list}} 
\begin{document}  
\large 
\begin{center} 
  {\bf   
PHOTO-INDUCED INTERMOLECULAR CHARGE TRANSFER\\ IN PORPHYRIN COMPLEXES} 
\vspace*{0.4cm}
 
 Michael Schreiber, Dmitry Kilin, and  
Ulrich Kleinekath\"ofer  
 
\vspace*{0.3cm} 
{\it Institut f\"ur Physik, Technische Universit\"at,  
D-09107 Chemnitz, Germany }   
\end{center} 
 

\textfloatsep 1ex
 \intextsep 7ex 
\baselineskip12pt 
 
 
\begin{center} 
\begin{minipage}[t]{12.7cm} 
\baselineskip12pt 
\renewcommand{\baselinestretch}{0.7} 
{\normalsize 
  Optical excitation of the sequential supermolecule $H_2P-ZnP-Q$ induces
  an electron transfer from the free-base porphyrin ($H_2P$) to the quinone
  ($Q$) via the zinc porphyrin ($ZnP$).  This process is modeled by
  equations of motion for the reduced density matrix which are solved
  numerically and approximately analytically. These two solutions agree
  very well in a great region of parameter space.  It is shown that for the
  majority of solvents the electron transfer occurs with the superexchange
  mechanism.
} 
\end{minipage} 
\end{center} 
\vspace*{0.1cm} 
 
\begin{center} 
{\bf  
I. INTRODUCTION} 
\end{center} 

\vspace*{-0.1cm} 
The investigation of  photoinduced charge transfer is important both for
the description of  natural photosynthesis \cite{j1} and for the
creation of artificial photoenergy-converting devices \cite{w1}.  For
experimental realizations of such artificial devices porphyrin complexes are
good candidates \cite{r4,z3,z4}.  Of major interest are those complexes
with an additional bridging block between donor and acceptor
\cite{j1,r4,z4,s1,d2}.

Electron transfer reactions can occur through different mechanism
\cite{s1,d2,s7}: sequential transfer (ST) or superexchange (SE).  Changing a
building block of the complex \cite{w1,r4} or changing the environment
\cite{z3} can modify which mechanism is most significant.  To clarify which
mechanism is present
one sequentially varies the energetics of the complex
\cite{j1,w1,r4,m1}.  This is done by radical substituting the porphyrin
complexes \cite{j1,w1,r4} or by changing the polarity of the solvent
\cite{r4,z3}.  Also the geometry and size of a bridging block can be varied and
in this way the length of the subsystem through which the electron has to be
transfered \cite{w1,d2,m6,m10}.
  
SE \cite{m6} occurs due to coherent mixing of the levels
\cite{r4,s1,s7,m6} and plays a role for any detuning of the energy levels
\cite{j1,w1,r4,z4}.  The transfer rate in this channel decreases
exponentially with increasing length of the bridge \cite{m6,m10}.
When incoherent effects such as dissipation and dephasing dominate
\cite{k10,k9}, the transfer is mainly sequential \cite{s1,m10}, i.~e.,
the  levels are occupied mainly in sequential order \cite{r4,s1}.
An increase in the bridge length induces only a small reduction in the
transfer rate \cite{s1,m10}.
  
In the case of coherent SE the dynamics is mainly
Hamiltonian and can be described on the basis of the Schr\"odinger equation.
The physically important results can be obtained by perturbation theory
\cite{m6}, most successfully by the Marcus theory \cite{m1}.  
In case
of ST the environmental influence has to be taken into
account.  
The more natural description of the relaxation process is based on
the density matrix (DM) formalism \cite{s1,d2,k9,l4,m11,k11,s8}.  The master
equation that governs the DM evolution as well as the
appropriate relaxation coefficients can be derived from such basic
information as system-environment coupling strength and spectral density
of the environment \cite{m11,k11,s8}.
  
The main physics of the system
can be described by
a DM equation which accounts for relaxation effects
phenomenologically\cite{d2,w4}.  
The master equation is analytically solvable only
for the simplest models \cite{k9,l4}.  
Most investigations are based on the numerical solution of this equation
\cite{s1,k10,m11}.  However, an estimations can be 
obtained within the
steady-state approximation \cite{d2}. Here we perform numerical
as well as approximate analytical calculations.

\begin{figure}[bt] 
\begin{minipage}{10cm}
\begin{center} 
\parbox{6.0cm}{\rule{-.01cm}{.1cm}\epsfxsize=8.0cm\epsfbox{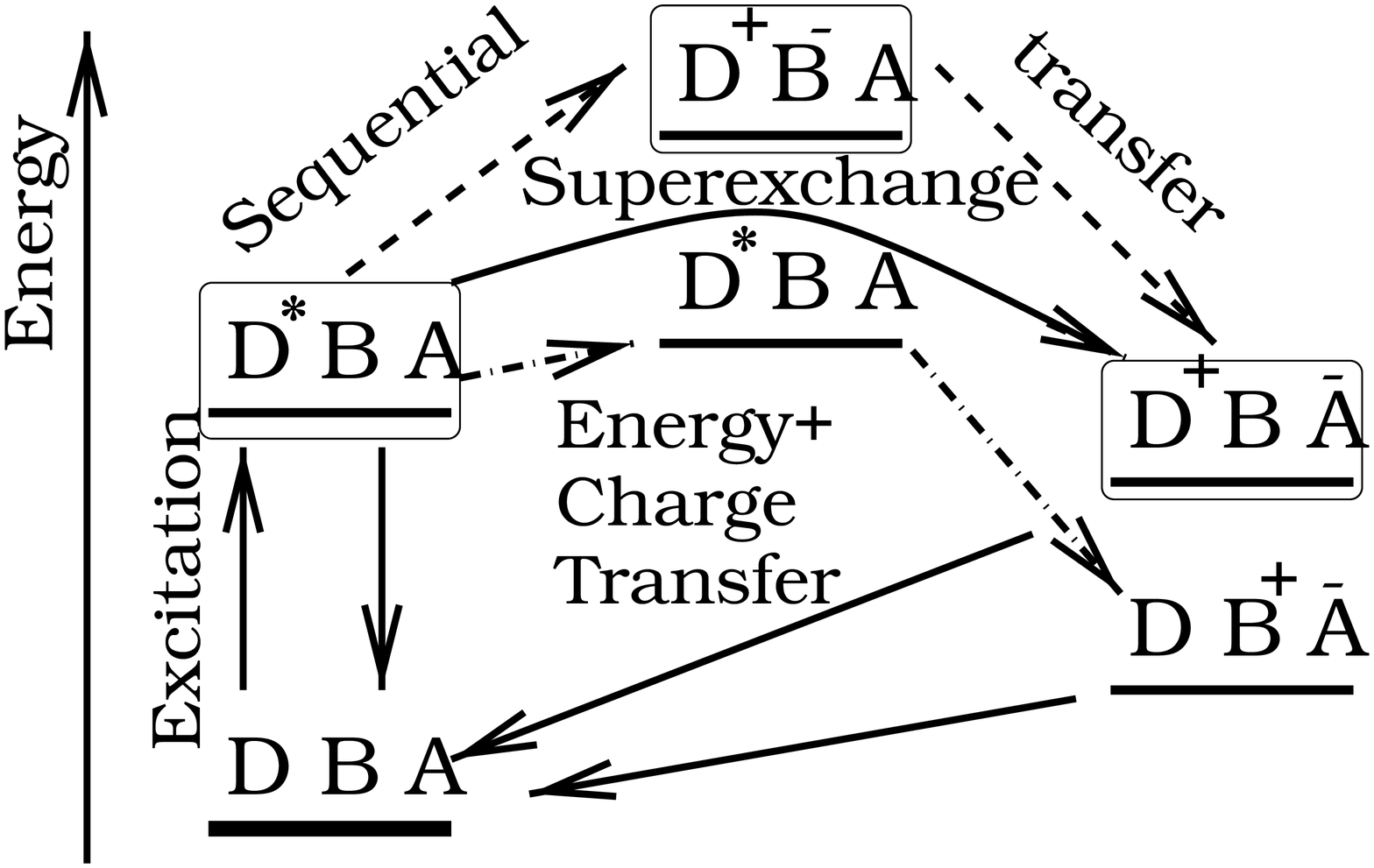}}
\end{center}
\end{minipage}\hspace*{0.3cm}\hfill
\begin{minipage}{3.8cm}  
\renewcommand{\baselinestretch}{1} 
\baselineskip12pt 
{\normalsize 
~~FIG.~1.
Schematic view of the energy levels in the $H_2P-ZnP-Q$ complex  
taken into account in calculation.   
The three states in the  boxes define   
the charge separation which   
can happen either by ST or by  SE.  
} 
\end{minipage} 
\end{figure} 
  
\begin{center} 
{\bf  
II. MODEL} 
\end{center}

\vspace*{-0.1cm}  
We investigate the photoinduced electron transfer in supermolecules that
consist of sequentially connected molecular blocks, namely donor, bridge,
and acceptor.  The donor (D) is not able to transfer its charge directly to
the acceptor (A) because of their spatial separation. D and A can exchange
their charges only through B (Fig. 1). In the present investigation the
supermolecule consists of free-base porphyrin($H_2P$) as donor, zinc
substituted porphyrin($ZnP$) as bridge, and benzoquinone as acceptor
\cite{r4}. In each of those molecular blocks we consider only two molecular
orbitals, the LUMO and the HOMO. Each of those orbitals can be occupied by
an electron ($|1\rangle$) or not ($|0\rangle$).  This model allows us to
describe the neutral nonexcited molecule
$|1\rangle_{HOMO}|0\rangle_{LUMO}$ and the following three states of
the molecule: neutral excited molecule $|0\rangle_{HOMO}|1\rangle_{LUMO}$,
positive ion $|0\rangle_{HOMO }|0\rangle_{LUMO}~,$  and negative ion
$|1\rangle_{HOMO}|1\rangle_{LUMO}$. Below  Roman indices indicate 
molecular orbitals ($m=0$ - HOMO, $m=1$ - LUMO), while  Greek indices
indicate  molecular blocks ($\mu=1$ - donor, $\mu=2$ - bridge, $\mu=3$ -
acceptor). Each of the electronic states has its
own vibrational substructure. However the time of vibrational relaxation
\cite{k11} is two orders of magnitude faster than the characteristic time
of the electron transfer \cite{r4}. Because of this we assume that only the
vibrational ground states play a dominant role in electron transfer.
  
One can describe the occupation of an orbital by an electron with the
appropriate creation operator 
$c^+_{\mu m}=|1\rangle_{\mu m}\langle0|_{\mu m}$ 
as well as its annihilation 
$c_{\mu m}=|0\rangle_{\mu m}\langle1|_{\mu m}$.  
Then $\hat n_{\mu}=\sum_m c^+_{\mu m}c_{\mu m}$ gives the number of
electrons in the molecular block $\mu$.
  
For the description of  charge transfer and other dynamical processes in 
the system we introduce the Hamiltonian  
\begin{eqnarray} 
  \label{1} 
\hat H=\hat H_S+\hat H_B+\hat  H_{SB}~, 
\end{eqnarray} 
 where  $H_S$ characterizes the supermolecule, $H_B$ the dissipative bath, and $H_{SB}$ the 
interaction between the two. $H_S$, however, 
includes  the static influence of the environment, namely of the 
solvent dipoles, which gives rise to a reduction of the energy 
levels, 
\begin{eqnarray} 
  \label{2} 
  \hat H_S=\sum_{\mu m}E_{\mu m} \hat n_{\mu m}+
\frac{3}{\epsilon_s+2}(\hat E_{el}+\hat E_{ion})+\hat V~, .
\end{eqnarray} 
The energies $E_{\mu
  m}$ are calculated in the independent particle approximation
\cite{kili98b}. $\epsilon_s$ denotes the static dielectric constant of the
solvent. 
$\hat E_{el} =\sum_{\mu}(\hat n_{\mu}-1){e^2}/({4\epsilon_0 r_\mu})$ 
describes the energy to create an
isolated ion.  This term depends on the characteristic radius $r_{\mu}$
of the molecular blocks.  
 $ \hat E_{ion}=\sum_{\mu} \sum_{\nu}
(\hat n_{\mu}-1)(\hat n_{\nu}-1){e^2}/({4\pi\epsilon_0}{r_{\mu \nu}})
$  includes the interaction
between the already created ions.
It depends on the distance between the molecular blocks $r_{\mu \nu}$.
The last contribution to the system Hamiltonian is the hopping term $\hat
V=\sum_{\mu \nu}v_{\mu \nu}(\hat V^+_{\mu \nu} + \hat V^-_{\mu \nu})((\hat
n_\mu-1)^2+(\hat n_\nu - 1)^2),$ which includes the coherent hopping
between each pair of LUMO $\hat V^-_{\mu\nu} = c^+_{\nu 1} c_{\mu 1}$, $\hat V^+ =
(\hat V^-)^+$ as well as the corresponding intensities $v_{\mu \nu}$. The
matrix elements of this operator give nonzero contribution only if one of
the states has a charge separation.  Because there is no direct connection
between donor and acceptor we assume $v_{13}=0$.

As usual the bath is given by harmonic oscillators with creation and
anhilation operators $a^+_\lambda$ and $a_\lambda$.  The system bath
interaction comprises both irradiative and radiative transitions. For $t
\ll 1-10~ns$ the latter one can be neglected .  The irradiative
contribution corresponds to energy transfer to the solvent and spreading of
energy over vibrational modes of the supermolecule
\begin{eqnarray} 
  \label{5} 
\hat H_{SB}=\sum_\lambda 
\sum_{\mu \nu} K_{\lambda, \mu \nu} v_{\mu \nu}(a^+_\lambda + 
a_\lambda)(\hat V^+_{\mu \nu}+\hat V^-_{\mu \nu})~,   
  \end{eqnarray} 
where $K_{\lambda, \mu \nu}$ reflects the interaction strength between bath mode $\lambda$ and 
quantum transition between LUMO levels of  molecules $\mu$ and $\nu$.  
  
Initially we use the whole density 
matrix of system and bath for the description of the dynamics.
 After applying the Markov and rotating wave approximations and tracing out
the bath modes \cite{s8} we obtain 
the equation of motion for the reduced density matrix (RDM)  
\begin{eqnarray} 
  \label{7} 
  \dot \sigma =-i/\hbar[\hat H_S, \sigma] &+&\sum_{\mu \nu}
  \Gamma_{\mu \nu} \{ (n(\omega_{\mu 1 \nu 1})+1)([\hat V^-_{\mu \nu}
  \sigma, \hat V^+_{\mu \nu}]+[\hat V^-_{\mu \nu}, \sigma \hat V^+_{\mu
    \nu}]) \nonumber \\ && + n(\omega_{\mu 1 \nu 1})([\hat V^+_{\mu \nu}
  \sigma, \hat V^-_{\mu \nu}]+[\hat V^+_{\mu \nu}, \sigma \hat V^-_{\mu
    \nu}])\}~,
\end{eqnarray} 
where the dissipation intensity $\Gamma_{\mu \nu}=\pi K^2_{\mu \nu}
\rho(\omega_{\mu 1 \nu 1}) v^2_{\mu \nu}$ depends on the coupling $K_{\mu
  \nu}$ of the transition $\mu 1 \leadsto \nu 1$ and on the bath mode of
the same frequency.  $\Gamma_{\mu \nu}$ depends also on the density $\rho$
of bath modes at the transition frequency $\omega_{\mu 1 \nu 1}$ and on the
corresponding coherent coupling $v_{\mu \nu}$ between the system states.
$n(\omega)$ denotes Bose-Einstein distribution.
 
For simplicity we introduce a superindex $i=\{ \mu m \}$,
the intensities of the dissipative 
transitions $d_{ij}=\Gamma_{ij}n(-\omega_{ij})$ 
between each pair of states, as well as the corresponding 
dephasing intensities $\gamma_{ij}=1/2\sum_k(d_{ik}+d_{kj})$. Taking these 
 simplifications into account one gets
\begin{eqnarray} \label{eq1}  
\dot \sigma_{ii} &=& -i/\hbar \sum_j  
(V_{ij}\sigma_{ji}-\sigma_{ij}V_{ji})  -\sum_i d_{ij}\sigma_{ii}+\sum_j  
d_{ji}\sigma_{jj}~,
\\  \label{eq2}  
\dot \sigma_{ij} &=& (-i\omega_{ij}-\gamma_{ij})\sigma_{ij}-i/\hbar  
V_{ij}(\sigma_{jj}-\sigma_{ii})~.  
\end{eqnarray}  
The simplification is that we do not calculate the system parameters, rather 
we extract them from experimental data. 
  
\vspace*{0.2cm}  
\begin{center} 
{\bf  
III. EXTRACTION OF SYSTEM PARAMETERS} 
\end{center} 

\vspace*{-0.1cm} 
The porphyrin absorption spectra \cite{r4} consist of  high frequency  
Soret bands and  low frequency $Q$ bands. In  case of $ZnP$ the $Q$ band  
has two subbands, $Q(0,0)$ and $Q(1,0)$. In the free-base porphyrin $H_2P$  
the reduction of symmetry 
induces a splitting of each subband  
into two, namely $Q^x(0,0)$, $Q^y(0,0)$ and $Q^x(1,0)$, $Q^y(1,0)$. So the  
emission spectra of $ZnP$ and $H_2P$ consist of two and four bands,  
respectively.  
Each of the abovementioned spectra can be represented as a sum of Lorentzians  
with  good precision. It is important to note that the spectra of  
porphyrin complexes contain all bands of the isolated porphyrins without  
essential changes.  
We use the lowest band of each spectrum. The corresponding frequencies  
and widths are shown in table\ \ref{tab:1}.

\begin{table}[t]  
  \begin{center}  
    \caption{Low-energy bands of the porphyrin spectra  for  
$CH_2Cl_2$ as solvent.}      \label{tab:1}  
      \begin{tabular}{p{2.7cm}p{2.7cm}p{2.7cm}p{2.7cm}p{2.4cm}}  
        \hline
 & \multicolumn{2}{c}{Absorption} & 
        \multicolumn{2}{c}{Emission} \\  & Frequency, eV & Width, 
        eV & Frequency, eV & Width, eV \\ \hline $H_2P$ & $\nu 
        _{00}^x=1.91$ & $\gamma _{00}^x=0.06$ & $\nu _{01}^x=1.73$ & 
        $\gamma _{01}^x=0.05$ \\ 
        $ZnP$ & $\nu _{00}=2.13$ & $\gamma 
        _{00}=0.07$ & $\nu _{01}=1.92$ & $\gamma _{01}=0.05$ \\
 \hline 
    \end{tabular}  
  \end{center}  
\vspace*{-0.3cm}
\end{table}
On the basis of the experimental spectra we determine $E_{D^*BA}=1.82eV$ and
$E_{DB^*A}=2.03eV$ (in $CH_2Cl_2$). The authors of Ref.\
\onlinecite{r4} give the energies of two other levels, 
$E_{D^+B^-A}=2.44eV$ and $E_{D^+BA^-}=1.42eV$. This allows to calculate
$E_{DB^+A^-}=1.21eV$.  
The 
hopping intensity
$v_{23}=v=2.2meV$ is calculated in
Ref.\ \onlinecite{w7}.  
On the other hand
Rempel et al.\ \cite{r4} estimate the electron coupling of the initially
excited and charged bridge states $v_{12}=V=65meV$.
We take the  intensity of the intermolecular conversions $\Gamma_{21}$,
$\Gamma_{23}$ in range $1-10\times 10^{11}s^{-1}$ 
\cite{kili98b}.
  
The main parameter which controls the electron transfer in a triad is the
relative energy of the state $D^+B^-A$. This state has a strong coupling to the
solvent that changes the energy of the state.  The
 values of the energy
$E_{D^+B^-A}$ calculated in the present model 
 are shown in table \ref{tab:2} for some solvents.
\begin{table}[b]  

  \begin{center}  
  \caption{Energy of the charged bridge state 
           and transfer rates in different solvents.}  
    \label{tab:2}  
\begin{tabular}{p{2.7cm}p{2.7cm}p{2.7cm}p{2.7cm}p{2.4cm}}  
\hline  
Solvent  & $75\%CH_2Cl_2$   $+25\%CH_3CN$ & $CH_2Cl_2$ & MTHF & CYCLO \\ 
 \hline  
$\epsilon _s$   & $15.75$ & $9.08$ & $6.24$ & $2.02$ \\ 
$\epsilon _\infty$  & $2.00$ & $2.01$ & $2.03$ & $2.03$ \\ 
$E_{D^+B^-A}$, $eV$  & $1.89$ & $2.86$ & $3.18$ & $5.30$ \\  
$k_{ET}$, $s^{-1}$  & $3.98 \times 10^{11}$ & 
$5.01 \times 10^{9}$ & $7.94 \times 10^{8}$ & $3.80 \times 10^8$ \\  
\hline  
     \end{tabular}  
  \end{center}  
\end{table}  
In table 2 $\epsilon_s$ denotes the static dielectric permittivity,  
$\epsilon_\infty$  the optic dielectric permittivity, MTHF  
2-methil-tetrahydrofuran, and CYCLO denotes cyclohexane. The calculated value  
$E_{D^+B^-A}=2.86eV$ deviates $15\%$  from the data of Ref.\  
\onlinecite{r4}.  
 
 \vspace*{0.2cm}  
\begin{center} 
{\bf  
IV. RESULTS} 
\end{center}

 \vspace*{-0.1cm} 
The time evolution of  charge transfer within  the
supermolecule is described by  Eqs.\ (\ref{eq1}) and (\ref{eq2}).
At initial time only the donor state is occupied.
 The calculations were 
performed with  two methods, direct numerical integration and
analytic approximation.
  
For the numerical simulation the eigenvalues and -vectors of the
system are calculated  and with these the time evolution
of the system is known. The  simulation of the system dynamics   
with the parameters determined in the previous section  
shows  exponential growth of the acceptor population.  
Such a behavior can be accurately  fitted to the formula  
$P_3(t)=P_3(\infty)[1-\exp{(-k_{ET}t)}],$  
where $k_{ET} \simeq 5 \times 10^9  s^{-1}$ and  $P_3(\infty)\simeq0.95$
for  $CH_2 Cl_2$ as solvent.  
The population  of the bridge state  does not exceed $0.005$.  
This shows that the SE mechanism dominates over the ST  
for the chosen set of parameters.  
In this case the system dynamics can be described by two values:
 the acceptor population   
at  infinite time $P_3(\infty)$ and the reaction rate $k_{ET}$  
that we deduce from the dynamics via  the following formula  
$k_{ET}=P_3(\infty)/\{\int_0^\infty [1-P_3(t)]dt\}$.  
  
The analytical approach is valid for the kinetic limit
$t \gg 1/\gamma_{ij}$.  In  Laplace-space we can replace
 $1/(i\omega_{ij}+\gamma_{ij}+s)$ by  $1/(i\omega_{ij}+\gamma_{ij})$,
 where $s$ denotes the Laplace
variable.  This  allows to simplify Eqs.\ (\ref{eq1}) and (\ref{eq2}) 
and  we define a new relaxation operator $(L\sigma)_{ii}^{new}=-\sum_i
g_{ij}\sigma_{ii}+\sum_j g_{ji}\sigma_{jj}.$ In this expression the
transition coefficients $g_{ij}$ contain both, dissipative and coherent
contributions
\begin{equation}  
g_{ij}=  
d_{ij}+v_{ij}v_{ji}\gamma_{ij}/[\hbar^2(\omega^2_{ij}+\gamma^2_{ij})].  
\end{equation}  
Assuming the bridge population to be zero  allows us to find  
the  dynamics of the acceptor state in the form  
$P_3(t)=P_3(\infty)[1-\exp{(-k_{ET}t)}],$   
where the final population $P_3(\infty)$ and the reaction rate $k_{ET}$  
are expressed in terms of the coefficients $g_{ij}$   
\begin{eqnarray}  
k_{ET}=g_{23}+\frac{g_{23}(g_{12}-g_{32})}{g_{21}+g_{23}}~,  
\hspace*{1cm}  
P_3(\infty)=\frac{g_{12}g_{23}}{g_{21}+g_{23}}(k_{ET})^{-1}.  
\end{eqnarray}

 \vspace*{0.2cm}  
\begin{center} 
{\bf  
V. DISCUSSION} 
\end{center}

 The following question will now be discussed: How does
the mechanism and speed of the reaction depend on a deviation of the
parameters from the determined values?
\begin{figure}[tb] 
\begin{center}
\hspace{.5cm} 
\rotate{
\rotate{
\rotate{
\epsfxsize=8.5cm\epsfbox{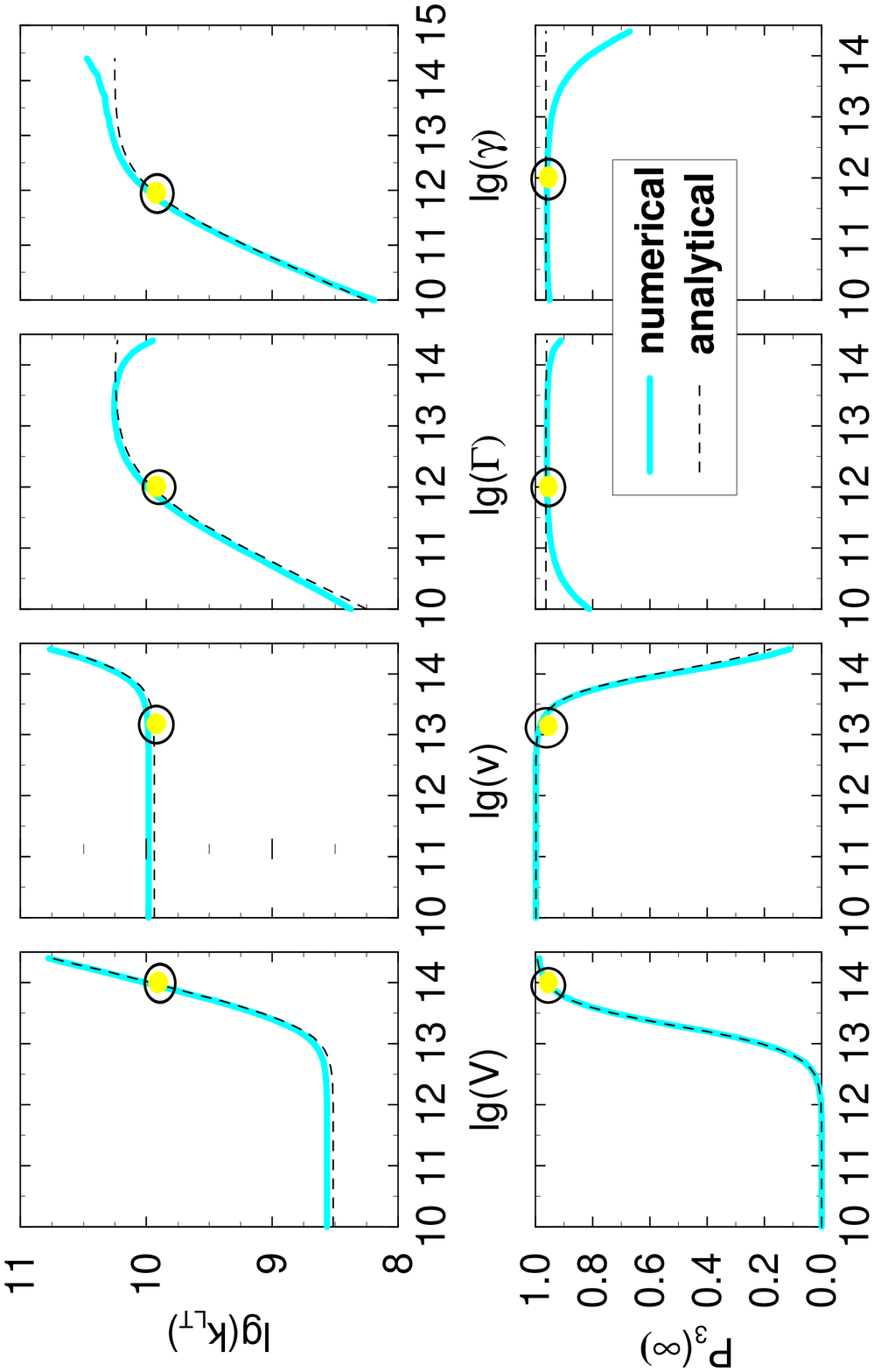}
}}}
\end{center} 
\vspace{-0.4cm}
\begin{minipage}{15.2cm}  
  \renewcommand{\baselinestretch}{1} \baselineskip12pt {\normalsize ~~FIG.~2.
    The dependence of the reaction rate (upper row) and final population of
    the acceptor state (lower row) on the  parameters $V=v_{12}$, $v=v_{23}$,
    $\Gamma=\Gamma_{21}$, $\gamma=\Gamma_{23}$.  Solid lines
    correspond to the numerical solution and dashed lines to the analytical 
solution.  The circles show the realistic parameter values for $CH_2Cl_2$ as solvent.}
\end{minipage} 
\end{figure} 
Namely which parameters have to be changed in order to change not only the
reaction rate quantitatively, but the dominant mechanism of reaction and
the qualitative behavior of dynamics at all.  To answer these questions we
calculate the system dynamics while varying one parameter at a time and
keeping the other parameters unchanged.  The dependencies of transfer rate
$k_{ET}$ and final population $P_3(\infty)$ on coherent couplings
$V=v_{12}$, $v=v_{23}$ and dissipation intensities $\Gamma=\Gamma_{21}$,
$\gamma=\Gamma_{23}$ are shown in Fig.\ 2.
 
\begin{figure}[tb] 
\hspace*{1.3cm}
\begin{minipage}{10.5cm} 
\begin{center}
\rotate{
\rotate{
\rotate{
\parbox{6.0cm}{\rule{-.01cm}{.1cm}\epsfxsize=6.0cm\epsfbox{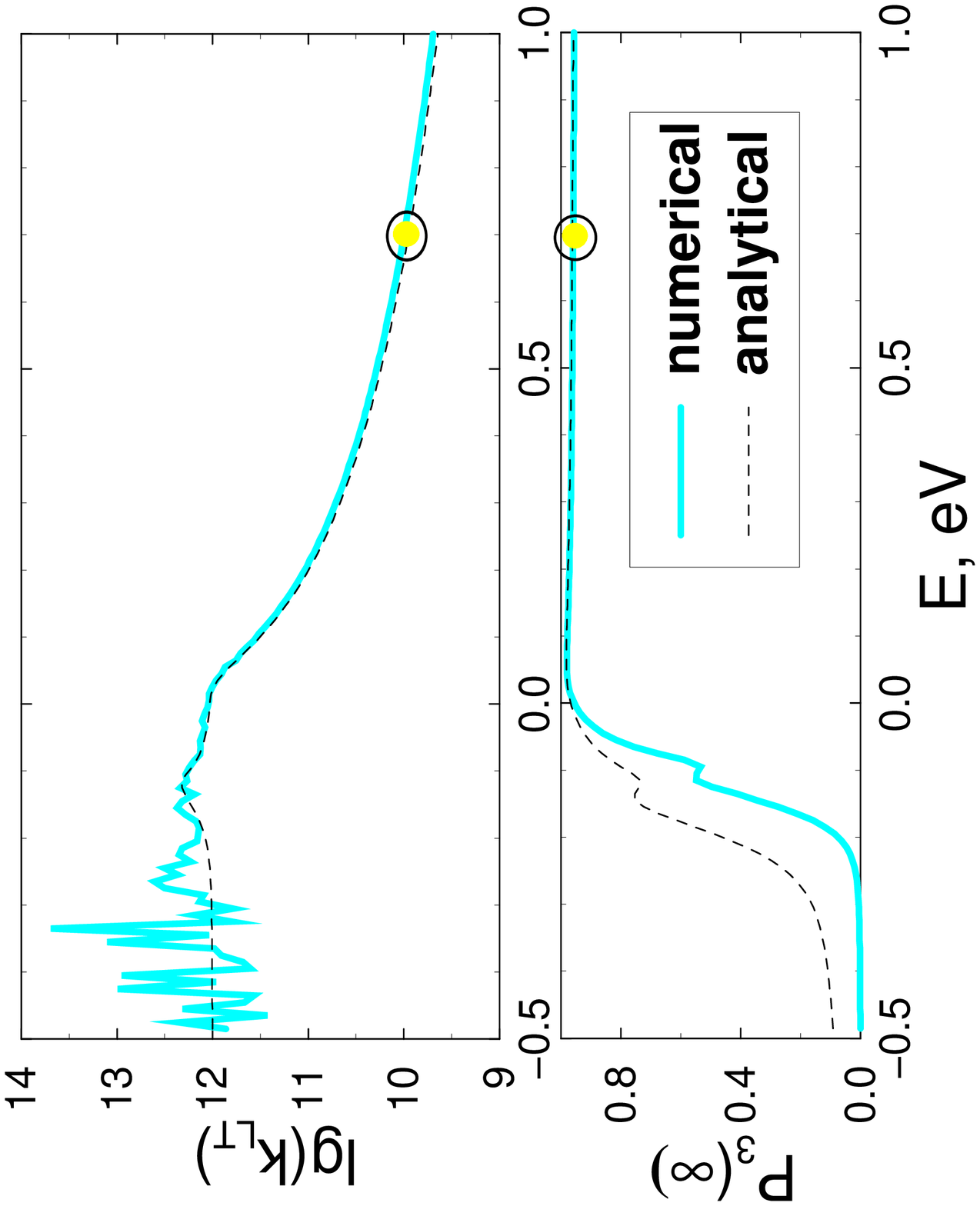}}
}}}
\end{center}
\end{minipage}\hspace*{-1.cm}\hfill 
\begin{minipage}{4.3cm}  
  \renewcommand{\baselinestretch}{1} \baselineskip12pt {\normalsize
    ~~FIG.~3.  Dependence of reaction rate (upper picture) and final
    acceptor population (lower picture) on the energy of the bridge state
    $E=E_{D^+B^-A}$.  Solid lines correspond to the numerical solution and
    dashed lines to the analytical solution.  }
\end{minipage} 
\vspace*{-1.cm}
\end{figure}

In particular, the decrease of the coherent coupling
$V$ induces a quadratic decrease of the reaction rate $k_{ET}$ until
saturation $V \sim 10^{10} ps^{-1}$.  Then $k_{ET}$ reaches its lower bound
and does not depend on $V$ anymore.  This corresponds to the crossover of
the reaction mechanism from SE mechanism to ST.  But, due to the big energy
difference between donor and  bridge state the efficiency of this ST
is extremely low, i.\ e.,\ $P_3 \leadsto 0$.  The considered variation of
the coherent coupling can be experimentally performed by exchanging
building blocks in the supermolecule.
  
The most crucial change in the reaction dynamics can be induced by changing the
energies of the system levels.  As discussed above this can be done by altering
the solvent.  Most important is the relative energy of the bridge state
$|D^+B^-A\rangle$. The results of the corresponding calculations are
presented in Fig.\ 3.  For high energies of the bridge state
$E_{D^+B^-A} \gg E_{D^*BA}$ the numerical and analytical results do not differ
from each other.  The reaction occurs with the SE mechanism that coincides
with the conclusion of Ref.\  \onlinecite{r4}.  
This is the case
for the most of solvents (see table 2). The smooth
decrease of energy induces an increase of the reaction rate up to the maximal
value near $1ps^{-1}$.

While the bridge energy approaches the energy of the donor state 
 the ST mechanism starts to contribute to the
process. As can be seen in table 2 this regime can be reached by the  use
strong polar solvents. 
The analytical solution does not coincide with the numerical one
anymore because the used approximations are no more valid in this region.
In the case  $E_{D^+B^-A}<E_{D^+BA^-}$  one cannot approximate the dynamics
of the acceptor population in the form $P_3 \sim
[1-\exp{(-k_{ET}t)}]$.  A high value of the bridge  energy ensures the
transition of the whole population to the acceptor state  $|D^+BA^-\rangle$. 
 In the intermediate case, when the bridge
state has the same energy as the acceptor state, the final population
spreads itself over these two states $P_3(\infty)=0.5$. 
At even lower bridge energies the population gets trapped at the bridge state.

We performed calculations for the electron transfer in the supermolecular
complex $H_2P-ZnP-Q$ within the RDM formalism.  The resulting analytical and
numerical reaction rates are in good agreement with each other and 
in qualitative correspondence with
experimental data \cite{j1,r4,kili98b}.  The SE mechanism of electron
transfer dominates over the sequential one.  The qualitative character of
the transfer reaction is stable with respect to a small variation of the
system parameter.  The crossover between the reaction mechanisms can be
forced by lowering the bridge state energy to the energy of the donor
state.
  
 \vspace*{0.2cm}  
\begin{center} 
{\bf  
REFERENCES} 
\end{center} 

\bibliographystyle{unsrt}

\end{document}